\documentclass[conference]{IEEEtran}
\usepackage[utf8]{inputenc}
\usepackage{graphicx,float,multirow}
\usepackage{amsmath}
\usepackage{bm}
\usepackage{amssymb}

\title{Massive Open Online Courses Temporal Profiling \\ for Dropout Prediction}
\author{\IEEEauthorblockN{Tom Rolandus Hagedoorn and Gerasimos Spanakis}
\IEEEauthorblockA{Department of Data Science and Knowledge Engineering\\
Maastricht University\\
6200MD, Netherlands\\
Email: t.rolandushagedoorn@student.maastrichtuniversity.nl,jerry.spanakis@maastrichtuniversity.nl}}

\begin{document}

\maketitle

\begin{abstract}\\
Massive Open Online Courses (MOOCs) are attracting the attention of people all over the world. Regardless the platform, numbers of registrants for online courses are impressive but in the same time, completion rates are disappointing. Understanding the mechanisms of dropping out based on the learner profile arises as a crucial task in MOOCs, since it will allow intervening at the right moment in order to assist the learner in completing the course. In this paper, the dropout behaviour of learners in a MOOC is thoroughly studied by first extracting features that describe the behavior of learners within the course and then by comparing three classifiers (Logistic Regression, Random Forest and AdaBoost) in two tasks: predicting which users will have dropped out by a certain week and predicting which users will drop out on a specific week. The former has showed to be considerably easier, with all three classifiers performing equally well. However, the accuracy for the second task is lower, and Logistic Regression tends to perform slightly better than the other two algorithms. We found that features that reflect an active attitude of the user towards the MOOC, such as submitting their assignment, posting on the Forum and filling their Profile, are strong indicators of persistence.
\end{abstract}


\section{Introduction}
Massive Open Online Courses (MOOCs), as the name suggests, are online courses open to anyone and aimed at teaching large audiences for free. The only entrance barrier to most of these courses is having access to a computer or a smart device with an internet connection. This, among other factors, causes them to have become increasingly popular over the last few years and attract hundreds or thousands of users in some cases. However, the number of students that effectively complete the courses is significantly smaller, as the dropout rate of many MOOCs is above 90\% \cite{khalil2014moocs}. Having  significant amounts of data collected on the users within a course lends itself well for quantitative statistical analysis with respect to the factors indicative of completion or dropout. 

This paper focuses on applying feature extraction and machine learning classification techniques on Maastricht University's MOOC on Problem Based Learning in order to better assess dropout behavior of learners. The contribution of this paper is threefold: First, a methodology of extracting features that describe user behavior within a MOOC (approach is applied to a specific dataset but can easily be extended to others as well) is described. Second, a machine learning framework (comparing three different algorithms, namely Logistic Regression, Random Forest and AdaBoost) to predict dropout within a temporal context (i.e. which is the exact time that a user will dropout) is presented. Finally, we conduct an analysis on which features are strong indicators of dropout (or persistence) from a MOOC. Novelty of this paper lies in the combination of static and temporal features while providing a straightforward machine learning framework for both predicting and highlighting indicators for dropouts.

The rest of the paper is organized as follows. Section \ref{sect:related} presents recent works relevant to dropout prediction. The dataset used in this paper is presented in Section \ref{sect:dataset}, followed by the methodology description in Section \ref{sect:tools}. Experimental setup and results are presented in Section \ref{sect:exp} and finally Section \ref{sect:discussion} concludes the paper.

\section{Related work}
\label{sect:related}
There have been several approaches tackling user dropout in MOOCs, but only a few take into account the temporal aspect. In \cite{balakrishnan2013predicting} authors utilized Hidden Markov Models to predict the persistence of a user in a MOOC, however their accuracy is not satisfying. Onah et. al \cite{onah2014dropout} delve into finding predictors of dropout by analyzing the behaviour and profile data of users. They conclude that the presence of interactions between users and tutors is a strong indicator of persistence, but their conclusions are not supported by any machine learning algorithm. Halawa et. al \cite{halawa2014dropout} explore many time related features, in order to find strong indicators of dropout, however their classification algorithm is very simple and could be improved. All these works attempt to predict one week ahead, except for \cite{ramesh2014learning} where they predict at three different time points during the course and \cite{taylor2014likely} where they provide predictions for all lags. Current work goes beyond \cite{taylor2014likely} by including profile information of learners so that prediction is also available when the course starts. Moreover, we compare three classification algorithms and come up with a way to measure feature importance.

\section{Dataset}
\label{sect:dataset}
\begin{figure}[h]
    \centering
    \scalebox{0.9}{
    \includegraphics[width = \linewidth]{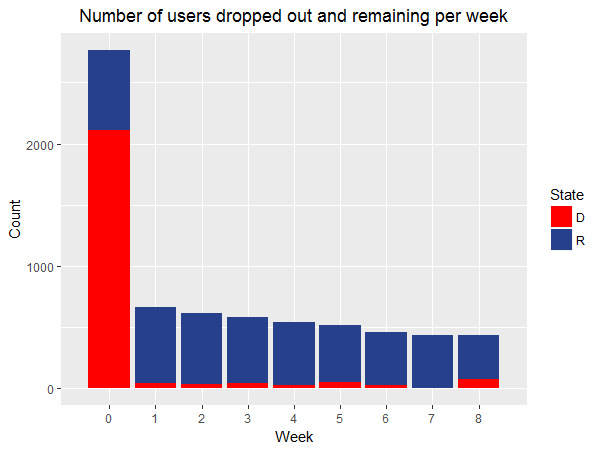}}
    \caption{Number of users dropped out and remaining per week, where R represents the remaining users and D the dropped out ones}
    \label{fig:dropouts}
\end{figure}

As mentioned previously, the dataset used in this paper is from Maastricht University's MOOC on Problem Based Learning that was offered from October till December of 2015. In total, after removing duplicates, there were 2769 users registered for the course, but only 358 of those completed it. This represents a dropout rate of 87\%, which corresponds to the rates found in similar literature \cite{taylor2014likely}. It is interesting to note that 75\% of the users failed to submit any assignment, and of those that submitted at least one assignment, 51\% completed the course.  Furthermore, it  out of the  37\% of  users that  left their profile empty, less than 1\%  completed the course.Figure \ref{fig:dropouts} shows the number of users that dropped out and those that remained per week.This graph shows a peak of dropouts in week 0, corresponding to users that did not start, and slightly  decreasing trend for the rests of the weeks. Furthermore, it shows that for most weeks the minority class, dropping out, is a rare event and rationalizes the use of specific techniques and metrics described in Sections \ref{sect:tools} and \ref{sect:exp} for dealing with such scarceness.
The dataset is composed of profile data, course activity data, forum data, video data and team data. 
All the features have to be dated as the classifiers should not use data posterior to the predicted week, hence team data and video data are omitted due to the lack of temporal information. 


\section{Methodology}
\label{sect:tools}
The dropout prediction framework consists of two main modules: the feature extraction and the prediction/classification algorithm. These sections describe in detail the process of constructing features from the data and the three classification algorithms tested.

\subsection{Feature extraction}
\begin{table}[]
\centering
\caption{Profile Data features}
\label{pdfeatures}
\scalebox{0.9}{
\begin{tabular}{|l|l|}
\hline
\textbf{Label} & \textbf{Definition {[}levels{]}}                  \\ \hline
\textit{pd\_a} & \textit{Country of residence {[}113{]}}           \\ \hline
\textit{pd\_b} & \textit{Primary language {[}62{]}}                \\ \hline
pd\_c          & Gender {[}4{]}                                    \\ \hline
pd\_d          & Biography                                         \\ \hline
pd\_e*         & Track chosen {[}3{]}                              \\ \hline
\textbf{pd\_f} & \textbf{Range of age of the user {[}5{]}}         \\ \hline
pd\_g*         & Why  choose this course? {[}4{]}                  \\ \hline
pd\_h*         & Role in education {[}5{]}                         \\ \hline
\textbf{pd\_i} & \textbf{Education experience {[}3{]}}             \\ \hline
\textbf{pd\_j} & \textbf{PBL experience {[}3{]}}                   \\ \hline
pd\_k*         & Areas of interest {[}6{]}                         \\ \hline
pd\_l*         & Work schedule preference {[}3{]}                  \\ \hline
pd\_m          & Time-zone {[}25{]}                                \\ \hline
\textbf{pd\_n} & \textbf{Anxious to discover the content? {[}5{]}} \\ \hline
\textbf{pd\_o} & \textbf{Determination to finish the MOOC {[}5{]}} \\ \hline
pd\_p          & Learning objectives                               \\ \hline
pd\_q          & First MOOC? {[}2{]}                               \\ \hline
pd\_r*         & Medium for finding this MOOC? {[}5{]}             \\ \hline
\end{tabular}}
\end{table}
The features extracted can be grouped into two different sets. The first set contains the Profile Data features, which are not fixed in time and thus the same ones are used for all the predictions, regardless of the week predicted. The second set consists of the Forum data features, the Course Activity data features and the Google Hangout features. These are all specific to each week, which means that to predict the dropouts for a certain week, only features of that week and of the weeks before it can be used. This set of features will be referred to as the temporal features. \\
The Profile data features shown in Table \ref{pdfeatures}  are all categorical except for biography and learning objectives where the values represent the standardized
length of the text written by the user. The features in \textbf{bold} are ordinal, which means that they are ordered, and conversely the rest of the categorical features which cannot be ordered, are said to be  nominal. The predictors followed by an * need some disambiguation:
\begin{itemize}
    \item pd\_e: the user has to choose between one of the following tracks:
    \begin{itemize}
        \item T1: Role of tutor in PBL
        \item T2: Designing PBL problems and courses
        \item T3: Assessment/organizational aspects of PBL
    \end{itemize}
    The track chosen will dictate which assignments must be handed in.
    \item pd\_g: The user can choose one of the following motivations:
    \begin{itemize}
        \item Personal interest
        \item Expand professional network
        \item Increase career opportunities
        \item other
    \end{itemize}
    \item pd\_h: The user can choose one of the following roles in education:
    \begin{itemize}
        \item Curriculum manager
        \item Not involved
        \item Teacher
        \item Educational adviser
        \item other
    \end{itemize}
    \item pd\_k: The user can choose one of the following areas of interest: 
    \begin{itemize}
        \item Arts - literature - philosophy
        \item Economics - Business - Trade
        \item Healthy body and healthy mind
        \item International relations - politics - law
        \item Science - Technology 
        \item None of these - No difference
    \end{itemize}
    \item pd\_l: The user can choose of the the following options for his preferred work schedule:
    \begin{itemize}
        \item Synchronously
        \item Asynchronously
        \item No preference
    \end{itemize}
    \item pd\_r: The user can choose one of the following mediums for finding the course:
    \begin{itemize}
        \item Professional network
        \item Social media 
        \item Maastricht University's website
        \item NovoEd website
        \item other
    \end{itemize}
\end{itemize}
The features in \textit{italics} have been dropped during early experimental stages as it was observed that they did not positively impact the predictions and that they slowed down the classifiers significantly. Furthermore, it is important to note that each categorical feature has one more level which corresponds to Not Communicated (NC) which means that the users left it blank.

\begin{table}[]
\centering
\caption{Temporal features}
\label{temporalfeatures}
\scalebox{0.9}{
\begin{tabular}{|l|l|}
\hline
\multicolumn{1}{|c|}{\textbf{ID}} & \multicolumn{1}{c|}{\textbf{DESCRIPTION}}                                                                                          \\ \hline
\textbf{f}                        & \textbf{Interaction with the forum}                                                                                                \\ \hline
fp                                & Number of forum posts                                                                                                              \\ \hline
fp\_l                             & Average forum post length                                                                                                          \\ \hline
fr\_ba                            & Number of replies received                                                                                                         \\ \hline
\textit{fr\_ba\_l}                & \textit{Average length of replies received}                                                                                        \\ \hline
\textit{fr\_ba\_u}                & \textit{Number of distinct user that replied}                                                                                      \\ \hline
fr\_ab                            & Number of replies given                                                                                                            \\ \hline
fr\_ab\_l                         & Average length of replies given                                                                                                    \\ \hline
\textit{fr\_ab\_p}                & \textit{Number of distinct posts replied to}                                                                                       \\ \hline
fc\_ba                            & Number of comments received                                                                                                        \\ \hline
\textit{fc\_ba\_l}                & \textit{Average length of comments received}                                                                                       \\ \hline
\textit{fc\_ba\_u}                & \textit{Number of distinct users that commented}                                                                                   \\ \hline
fc\_ab                            & Number of comments given                                                                                                           \\ \hline
fc\_ab\_l                         & Average length of comments given                                                                                                   \\ \hline
fc\_ab\_p                         & Number of distinct posts commented on                                                                                              \\ \hline
\textit{{[}...{]}\_i*}            & \textit{\begin{tabular}[c]{@{}l@{}}All of the above containing "\_ba" but\\ where the user replying is an instructor\end{tabular}} \\ \hline
\textbf{a}                        & \textbf{Assignment submitted}                                                                                                      \\ \hline
\textbf{ar}                       & \textbf{Assignment review submitted}                                                                                               \\ \hline
\textbf{gh}                       & \textbf{Participated at Google Hangout session}                                                                                    \\ \hline
\end{tabular}}
\end{table}

Table \ref{temporalfeatures} shows the temporal features. All are numerical and standardized, except for the ones in \textbf{bold} which are Boolean. The features in \textit{italics} have been dropped for the same reasons as in Table \ref{pdfeatures}.
Furthermore, "\textit{[...]\_i}" represents the 6 features containing "\_ab" but where the user replying is an instructor. From the literature in Section \ref{sect:related}, it was expected that these features would perform well. However, the instances in which these features take positive values are very rare, which leads to them decreasing the performance of the classifiers, hence they have been dropped.

\subsection{Logistic Regression}
Logistic Regression is a classification technique 
widely used  for predicting the outcome of binary classes \cite{cox1958regression}. 
Given a training set of $N$ instances, let $\mathbf{x_i}$ be the feature vector of instance $i = 1,2,...,N$ and be of length $K +1$ for the $K$ features and $x_{i0} = 1$ as a dummy variable for $\beta_0$.
Let $\mathbf{y}$ be the column vector of length $N$ representing the binary class of each instance.$\bm{\beta}$ is the column vector of coefficients of the predictors computed by the logistic regression. The probability that a given instance is predicted as a success is calculated by the logistic function defined as follows:
\begin{equation}
    \pi_i = Pr(y_i = 1|\bm{x_i}) = \frac{1}{1+e^{-(\bm{\beta x_i})}}
\end{equation}
Hence, the probability that an outcome is a failure is equal to  $1-\pi_i$.

\subsubsection{Training}
Maximum Likelihood Estimation (MLE) is used to iteratively train the logistic regression. It estimates the values of  $\bm{\beta}$ such as to maximize the log-likelihood function defined as follows:
\begin{equation} \label{log-likelihood}
   \ell(\bm{\beta}) = \sum_{i=1}^N y_i \textrm{log}(\pi_i)+\sum_{i=1}^N (1-y_i) \textrm{log}(1-\pi_i)
\end{equation}
To estimate $\bm{\beta}$, it initiates with a set of random coefficients, then at each iteration, it uses Newton's method to find the steepest gradient between the current predictions and the actual classes, and updates the coefficients of the features accordingly. It  repeats the process until convergence of the coefficients \cite{menard2002applied}.
\subsubsection{Feature selection}
The significant number of features extracted justifies the implementation of a feature selection method to improve the model.
The most important incentives for using such techniques are decreasing the variance of the coefficients, decreasing over-fitting, and increasing model interpretability \cite{janecek2008relationship}. 

The feature selection method applied to the Logistic Regression   is  the Elastic-net  regularization. It is a weighted combination of two other regularization techniques, namely the Lasso and Ridge regressions. The idea is to minimize the penalized negative log-likelihood ($PNLL$) function  which is defined as follows: 
\begin{equation} \label{PNLL}
   PNLL = -\ell(\bm{\beta}) + \lambda J(\bm{\beta})
\end{equation}
Where $-\ell(\bm{\beta})$ is obtained from Eq. \ref{log-likelihood},  $J(\bm{\beta})$ is the shrinkage penalty  and $\lambda \geq 0$ is a tuning parameter which controls how much weight is given to each term. The Elastic-net shrinkage penalty is defined as follows:
\begin{equation}
    J^{EN}(\bm{\beta}) = (1-\alpha) \sum_{k=0}^K \frac{\beta_k^2}{2} + \alpha \sum_{k=0}^K |\beta_k| 
\end{equation}
Where $\alpha \in [0:1]$ is a tuning parameter controlling the weight applied to each norm. When $\alpha = 0$,  a ridge regularization is performed, whereas when $\alpha = 1$ a lasso regularization is performed. H. Zou et. al\cite{zou2005regularization}, who first proposed the elastic-net approach, argue that it inherits the benefits of the Lasso method, while at the same time being able to group strongly correlated predictors. Both $\lambda$ and $\alpha$ are  tuned through cross-validation, as will be explained in Section \ref{section:trainingValidation}.

\subsection{Random Forest}
Random Forest is an ensemble learning method
often used for classification and that has shown to perform  well \cite{svetnik2003random}. The main idea of this classifier is to create many decision trees independently, through a technique called bagging, and combining their outputs in order to make predictions. To select the best predictors, the decision trees use the Gini index defined as follows:  
\begin{equation}
    I_G(f) = 1- \sum_{i=1}^J f_i^2
\end{equation}
Where $f_i$ is the fraction of instances of class $i$ and $J$ is the number of classes. The lower the Gini index of a predictor, the better it is at splitting the data. 
\subsubsection{Bagging}
In order to overcome the fact that decision trees are prone to overfitting the data, Random Forest uses a specific implementation of a  technique called bootstrap aggregating (bagging) to grow many trees independently and combine their results \cite{pal2005random}. To do so, it samples with replacement $B$ times from the training set to generate $B$ new samples of equal size to the original one, and trains a classification tree on each sample using  $\sqrt{p}$ randomly selected predictors of the total $p$ predictors available. Once all the trees have been trained, the Random Forest will be able to make predictions for unseen data by taking the mode of the predictions of each tree .\\

\subsection{AdaBoost}
AdaBoost,  akin to Random Forest, is an ensemble learning algorithm that uses decision trees as weak learners in order to obtain a strong learner. The main idea of this classifier is to build decision trees sequentially through a technique called boosting, such that each tree improves on the previous one. AdaBoost is considered to be the first successful implementation of a boosting algorithm, and is still considered to be a very good classifier for binary classes \cite{freund1995desicion}. The decision trees used in this classifier are similar to those used in Random Forest, the only difference being that they generally only have a depth of a few nodes.

\subsubsection{Boosting}
Boosting, in the context of decision trees, consists of growing a tree on the training dataset where all the samples have the same weights, then re-weighting the dataset such that the weights of misclassified samples increase and  repeating the process for an arbitrary number of trees. Once all the trees have been built, the final classifier combines the predictions through a weighted vote approach, where each weight is a function of the individual tree's performance measure. 
\subsubsection{Training}
Let $\bm{Z}$ be the training set such that $\bm{z_n}=(\bm{x_n},y_n) \textrm{ for } n = 1,...,N$ where $\bm{x_n}$ is the predictor vector and $y_n \in \{-1,+1\}$ is the class label of instance $\bm{z_n}$. Then, let $\bm{W}$ be the  vector of weights associated to each training instance and initiated such that $\bm{W}(\bm{z_n}) = \frac{1}{N} , \forall n \in N$. Now, for $t = 1,...,T$ with $T$ being an arbitrary number of trees, let $\bm{S_t}$ be the training set obtained by sampling with replacement $N$ times from $\bm{Z}$ with respect to the weights $\bm{W}$ and let $h_t(\bm{S_t})$ be the tree trained on that dataset. The error of the weak learner is  calculated as follows: 
\begin{equation}
    \epsilon_t = \sum_{i:h_t(x_s)\neq y_s} \bm{W}(\bm{z_s}), \quad \forall \bm{z_s} \in \bm{S_t}
\end{equation}
The weights of the  misclassified instances are then increased as follows:
\begin{equation}
    \bm{W}(\bm{z_s}) = \bm{W}(\bm{z_s}) \times e^{\alpha_t \times I(y_s \neq h_t(\bm{x_s}))}, \quad \forall \bm{z_s} \in \bm{S_t}
\end{equation}
Where $I(y_s \neq h_t(\bm{x_s}))$ is an indicator function that takes the value 1 if the instance was misclassified and 0 otherwise. And the parameter $\alpha_t$ is defined as such: 
\begin{equation}
    \alpha_t = \frac{1}{2}\bigg(\frac{1-\epsilon_t}{\epsilon_t}\bigg) 
\end{equation}
Then, $\bm{W}$ is normalized in order to represent a distribution function, and the steps of sampling, training and re-weighting are repeated until $T$ decisions trees have been generated. Once all the trees have been trained, the AdaBoost classifier is obtained by weighted majority voting defined as follows:
\begin{equation}
    AdaBoost(\bm{x}) = \textrm{sign}\bigg(\sum_{t=1}^T \alpha_t h_t(\bm{x})\bigg)
\end{equation}

\section{Experimental setup}
\label{sect:exp}
\subsection{Training-validation}\label{section:trainingValidation}
\subsubsection{k-fold cross validation}
K-fold cross validation (k-CV) is used in order to train and validate the models. To do so, it partitions the data into $k$ equal sized samples (folds), then $k-1$ of those are used to train the model and the remaining one is used to test the model. This process of training and testing is repeated $k$ times, using a different testing sample every time. The performance metrics of the $k$ iterations are then averaged and used as an estimate of the performance of the model. It has been shown that validating through k-CV when one of the classes is relatively rare is generally one of the best options \cite{kohavi1995study}. The classifiers discussed in this paper include hyperparameters that are tuned through k-CV. To avoid ambiguity, the k-CV for evaluating the models will be referred to as the outer loop and the one for tuning the hyperparameters as the inner loop. This naming reflects the fact that the inner loop k-CV is applied to each training set created in the outer loop.
\subsubsection{SMOTE}
Synthetic Minority Over-sampling Technique (SMOTE) is a method regularly used to improve the performance of classifiers on datasets with rare events \cite{he2009learning}. To do so it synthesizes new samples of the minority class, dropouts in our case, using k-Nearest-Neighbours. SMOTE is applied to every training set created in the outer loop of the nested k-CV, but never to the test sets, as these must reflect the real distribution of the data. 

\subsection{Evaluation metrics}
The choice of performance metrics is important as different metrics assess different aspects of the models, and using an inadequate one can be misleading. For instance, accuracy, one of the most commonly used metrics for binary classification tasks, is shown to be ineffective at representing the performance of a model when the data set is  imbalanced \cite{weiss2004mining}. 
Table \ref{table:confusion} represents a confusion matrix, which compares the predicted classes to the reference values.

\begin{table}[]
\centering
\caption{Confusion Matrix}
\label{table:confusion}
\begin{tabular}{c|c|c|c|}
\cline{3-4}
\multicolumn{2}{c|}{\multirow{2}{*}{}}                                                                   & \multicolumn{2}{c|}{Predicted} \\ \cline{3-4} 
\multicolumn{2}{c|}{}                                                                                    & Pos            & Neg           \\ \hline
\multicolumn{1}{|c|}{\multirow{2}{*}{\begin{tabular}[c]{@{}c@{}}Actual\\  Class\end{tabular}}} & Pos (P) & TP             & FN            \\ \cline{2-4} 
\multicolumn{1}{|c|}{}                                                                         & Neg (N) & FP             & TN            \\ \hline
\end{tabular}
\end{table}


\subsubsection{AUROC}
The Area Under the Receiver Operating Characteristic curve (AUROC) is a newer metric than the aforementioned accuracy and has become the norm in measuring the performance of binary classifiers \cite{huang2005using}. It takes advantage of the fact that most classifiers output probabilities instead of binary classes. To do so, it calculates the relationship between the True Positive Rate (TPR or Sensitivity or Recall) and the False Positive Rate (FPR or Fall-out) while sweeping through  threshold values for the output probabilities. The TPR and FPR are calculated as follows: 
\begin{equation}
    TPR = \frac{TP}{TP+FN}
\end{equation}
\begin{equation}
    FPR = \frac{FP}{FP+TN}
\end{equation}

The ROC curve is then obtained by plotting the values calculated with the different thresholds. Using this curve, the AUROC is derived by calculating the area underneath it. This one number metric summarizes the goodness of fit of a classifier. J. Davis et. al\cite{davis2006relationship} argue that the AUROC can be overly optimistic if the dataset is highly imbalanced, and propose to use other metrics such as the Area Under the Precision-Recall Curve (AUPRC). However, in our scenario, the cost of missing dropouts is considerably higher than the cost of predicting too many dropouts. Hence, AUROC is considered to be a more suitable metric.

\subsubsection{F2 measure}
The AUROC measures the performance of the classifiers over all the thresholds of the outcome probabilities, which is a good measure for comparing different models within the experiments. But in order to measure how well the classifiers would perform in a real life scenario where hard class labels have to be predicted, we use the F2 measure. This measure is a variant of the $F_\beta$ measure, which is defined as follows:
\begin{equation}
    F_\beta = (1 +\beta^2) \frac{\textrm{precision}\times \textrm{recall}}{(\beta^2 \times \textrm{precision})+\textrm{recall}}
\end{equation}
where $\beta = 2$, and precision is calculated as follows:
\begin{equation}
\textrm{precision} = \frac{\sum TP}{\sum TP + FP}
\end{equation}

\subsection{Experimental Results}
Throughout the experiments we hold a few parameters fixed in order to be able to compare the results . Each experiment uses the  k-CV approach where $k= 10$ for the outer loop and $k=3$ for the inner loop. The reason for choosing a relatively small $k$ value for the inner loop is because using greater ones would result in having the test sets of the inner folds contain too few positive cases, especially when predicting the exact week of dropout. 
\subsubsection{State prediction}
\begin{figure}[h]
    \centering
    \includegraphics[width = \linewidth, height = 6cm]{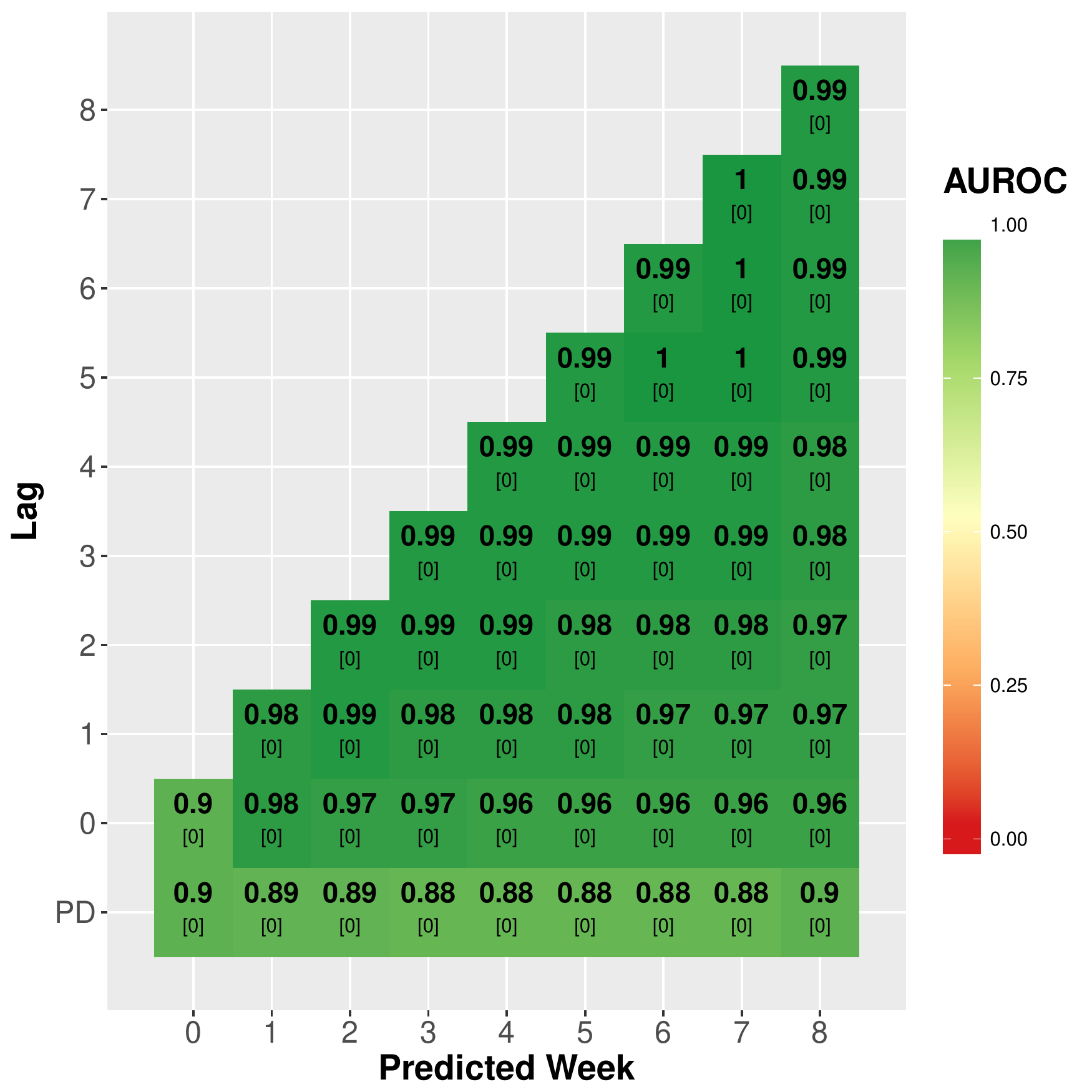}
    \caption{AUROC measures for state prediction with Logistic Regression}
    \label{fig:lr_notExactWeek_auroc}
\end{figure}

\begin{figure}[h]
    \centering
    \includegraphics[width = \linewidth, height = 6cm]{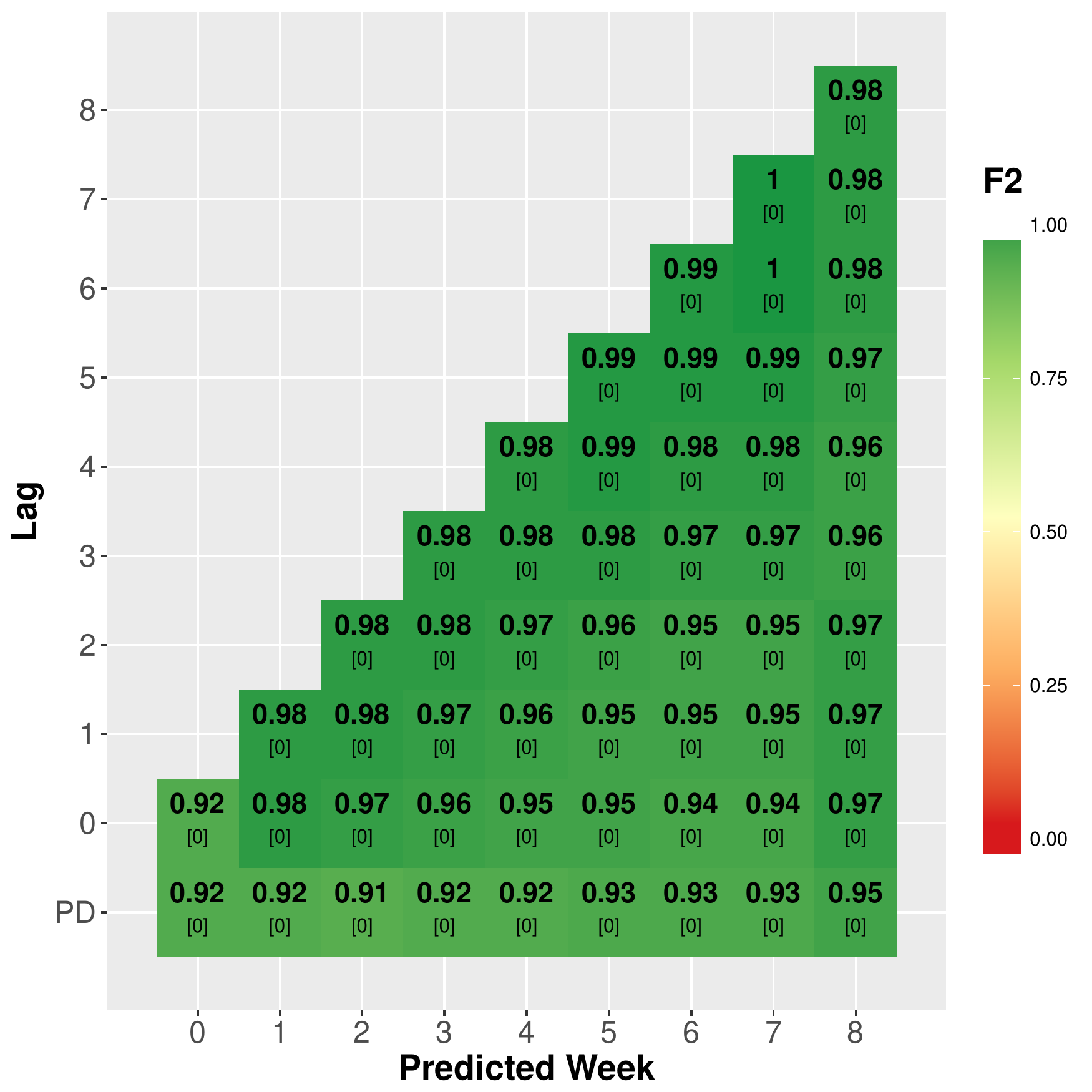}
    \caption{F2 measures for state prediction with Logistic Regression}
    \label{fig:lr_notExactWeek_f2}
\end{figure}

The first experiments are aimed at predicting the state of the users, which comes down to predicting which users will still participate from a certain week on. Throughout this paper we described  the dataset as being unbalanced such that the dropouts are the minority class, however this is reversed when predicting the states. Indeed, all users that have dropped out already are still used for future predictions, which means that for week 8 the remaining users represent 13\% of the data. Figures \ref{fig:lr_notExactWeek_auroc} and \ref{fig:lr_notExactWeek_f2} show the AUROC and F2 measures of the Logistic Regression trained and tested on every lag/week combination. We omitted the graphs corresponding to the same experiments using Random Forest and AdaBoost because the values differed at most by 0.01 for a few lag/week combinations. We can see that both the AUROC and the F2 measures are high throughout the combinations, except for the predictions using only the Profile features and for those of week 0 with a lag of 0. This shows that all three algorithms perform very well for predicting whether a user has dropped out by a certain week.
\subsubsection{Dropouts per week}
\begin{figure}[h]
    \centering
    \includegraphics[width = \linewidth, height = 6cm]{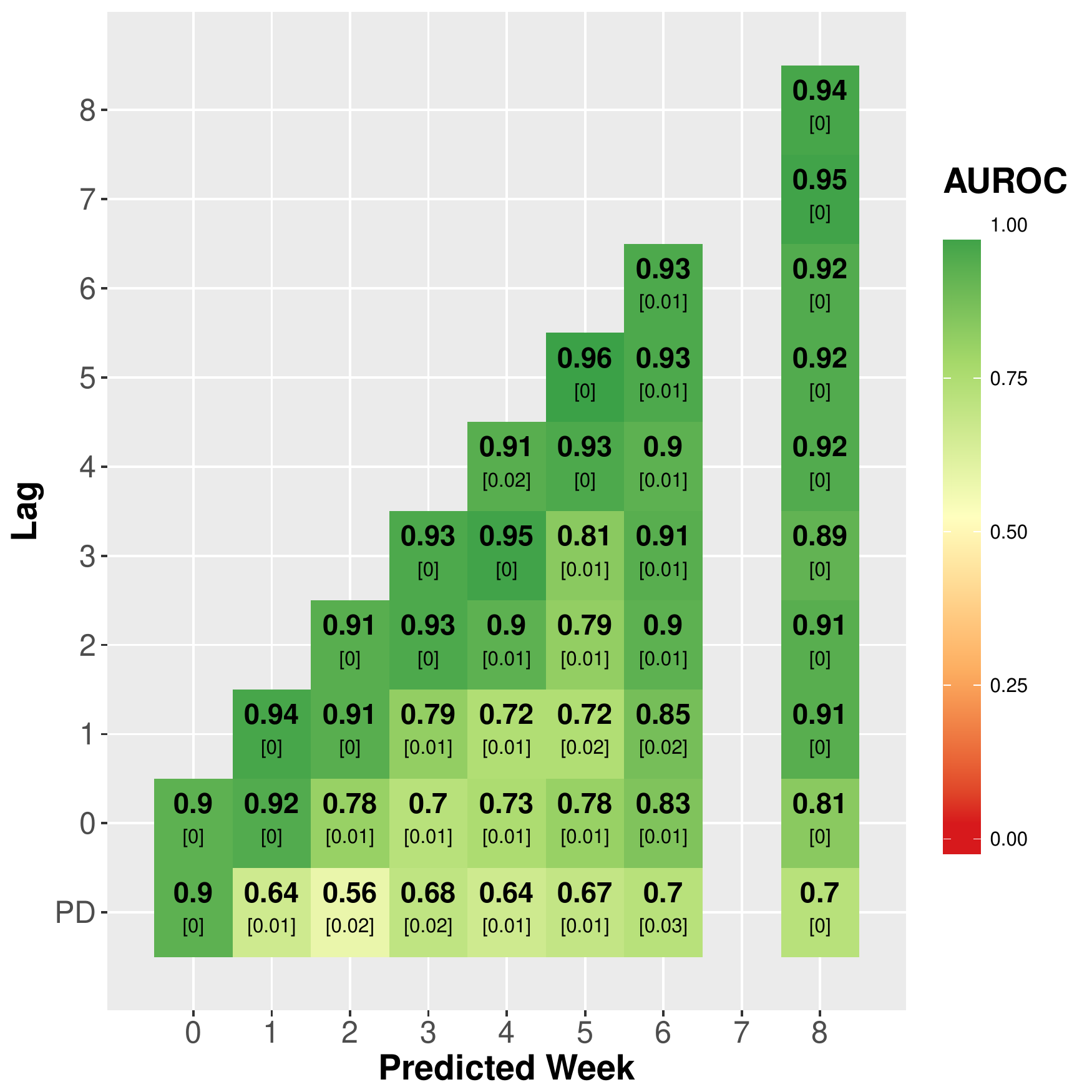}
    \caption{AUROC measures for dropout week prediction with Logistic Regression}
    \label{fig:lr_exactWeek_auroc}
\end{figure}

\begin{figure}[h]
    \centering
    \includegraphics[width = \linewidth, height = 6cm]{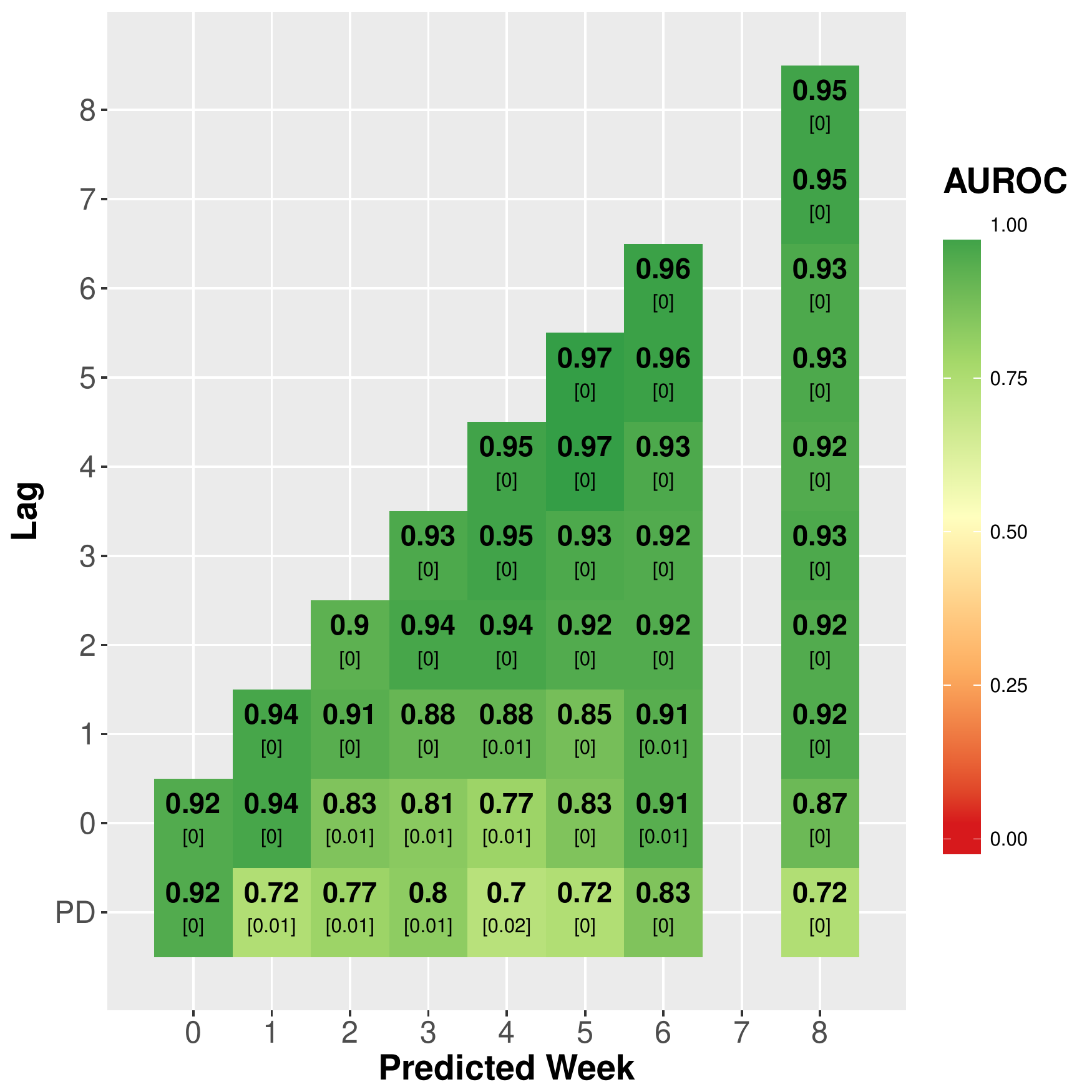}
    \caption{AUROC measures for dropout week prediction with AdaBoost}
    \label{fig:gbm_exactWeek_auroc}
\end{figure}

\begin{figure}[h]
    \centering
    \includegraphics[width = \linewidth, height = 6cm]{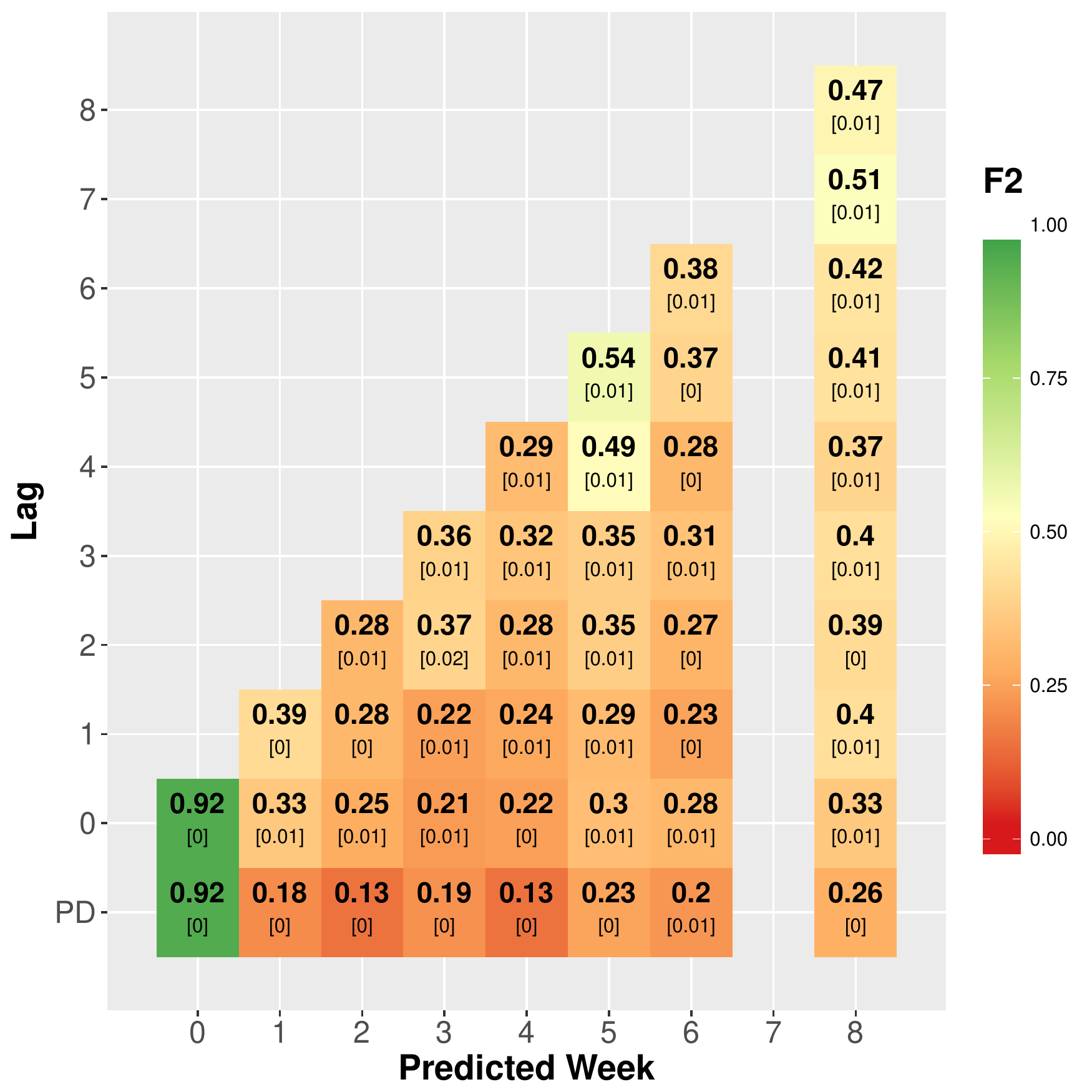}
    \caption{F2 measures for dropout week prediction with Logistic Regression}
    \label{fig:lr_exactWeek_f2}
\end{figure}

\begin{figure}[h]
    \centering
    \includegraphics[width = \linewidth, height = 6cm]{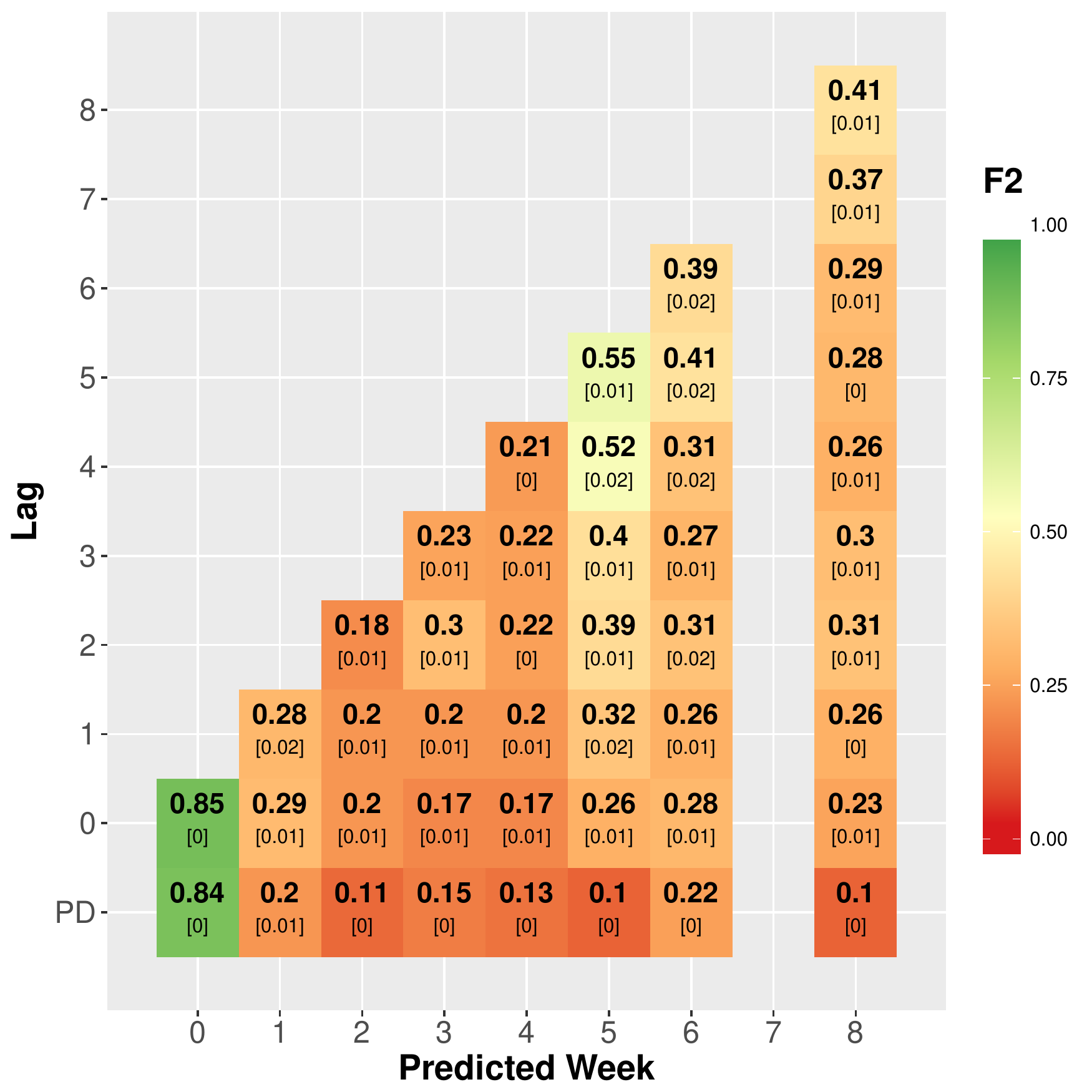}
    \caption{F2 measures for dropout week prediction with Random Forest}
    \label{fig:rf_exactWeek_f2}
\end{figure}

\begin{figure}[h]
    \centering
    \includegraphics[width = \linewidth, height = 6cm]{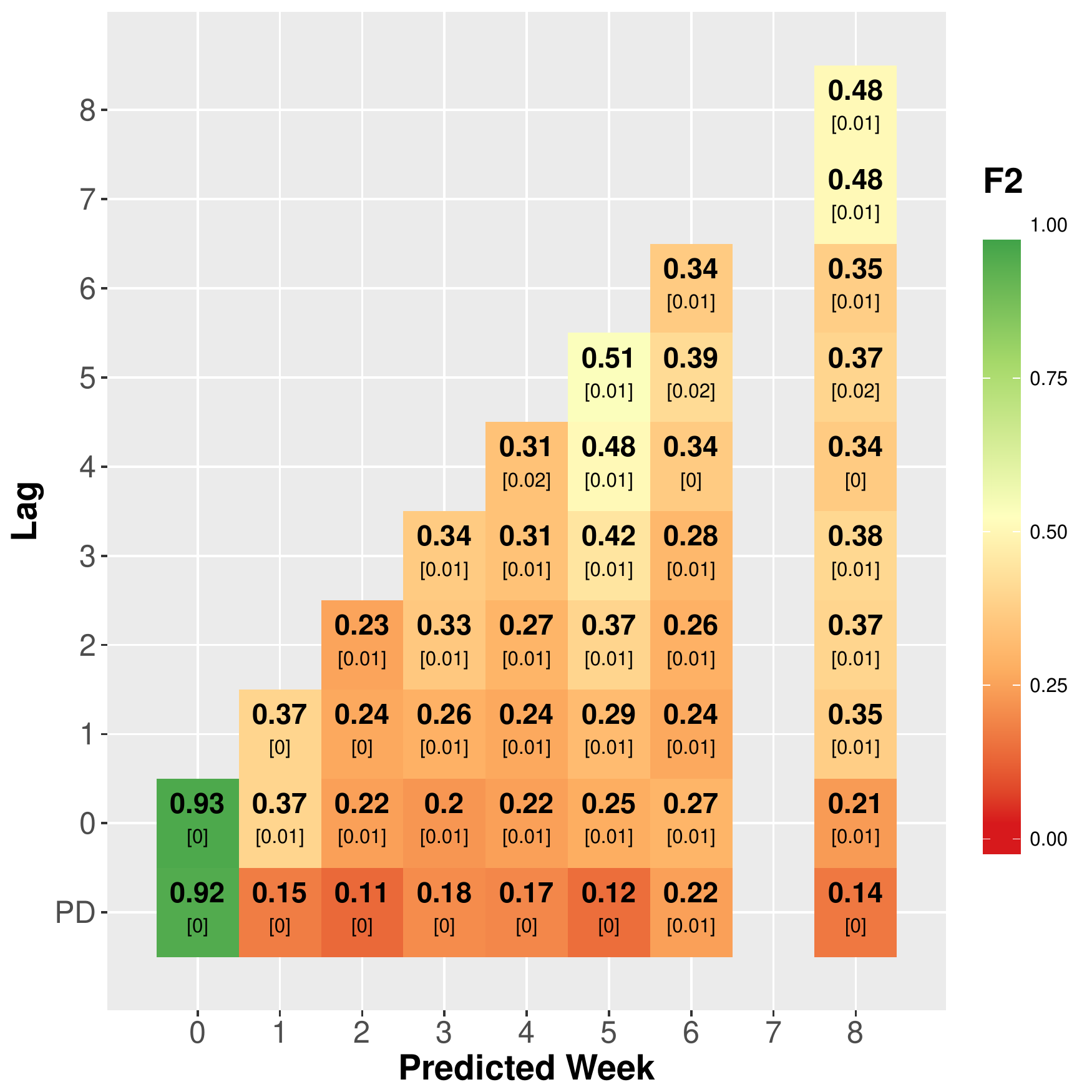}
    \caption{F2 measures for dropout week prediction with AdaBoost}
    \label{fig:gbm_exactWeek_f2}
\end{figure}

\
The second set of experiments focuses on predicting dropouts for each specific week. In this scenario, the minority class becomes the dropouts, except for week 0 as can be seen in figure \ref{fig:dropouts}.
Furthermore, in order to overcome the fact that the events of the minority class appear only rarely in most of the weeks, we use the SMOTE technique to over-sample the dropouts  in the training sets. The Figures \ref{fig:lr_exactWeek_auroc},\ref{fig:gbm_exactWeek_auroc},\ref{fig:lr_exactWeek_f2},\ref{fig:rf_exactWeek_f2} and \ref{fig:gbm_exactWeek_f2} show the AUROC and the F2 measures for each of the classifiers. We directly see that these values are lower than for the previous experiments, which was expected. Week 7 has no values because no user dropped out during that week. Furthermore, we notice an important difference between the two  metrics. This is mainly due to the fact that the AUROC  barely penalizes for predicting too many dropouts, whereas the F2 measure highly penalizes this. Hence, the low F2 measures reflect low precision values, but the recall values are still relatively high. In other words, these algorithms tend to predict too many dropouts which drastically lowers the F2 score, however their performance at identifying those that will actually drop out, which is captured in the recall, is still quite good. Furthermore, the values for week 0 are still very high, this is because many learners drop out that week, hence the algorithms do not considerably overestimate the number of dropouts. We can see that overall, AdaBoost slightly outperforms the two other classifiers with respect to AUROC, while Logistic Regression slightly outperforms the two others with respect to the F2 measure.
\begin{figure}[h]
    \centering
    \includegraphics[width = \linewidth, height = 6cm]{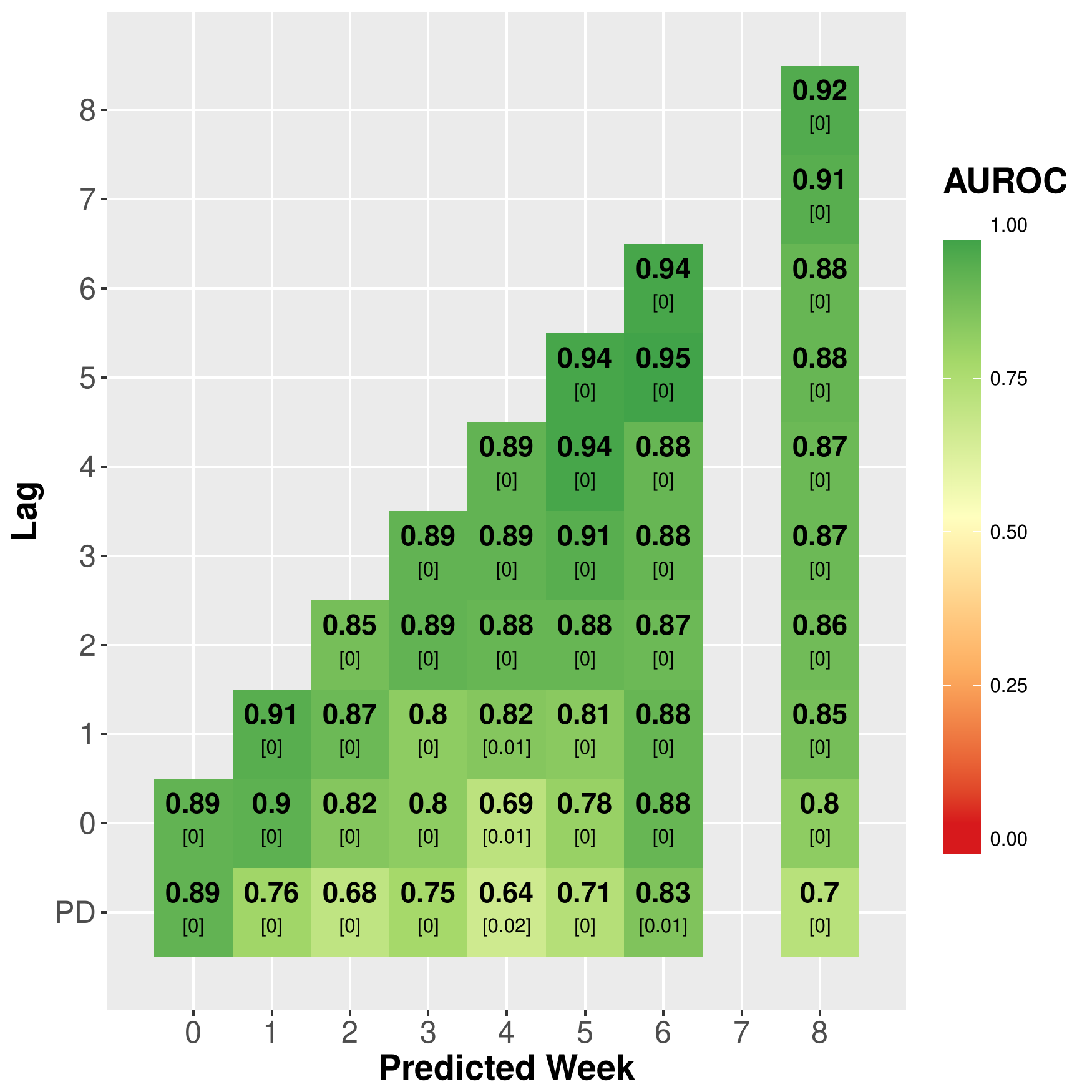}
    \caption{AUROC measures for dropout week prediction with Random Forest}
    \label{fig:rf_exactWeek_auroc}
\end{figure}

\subsubsection{Feature importance}
\begin{figure}[h]
    \centering
    \includegraphics[width = \linewidth, height = 10cm]{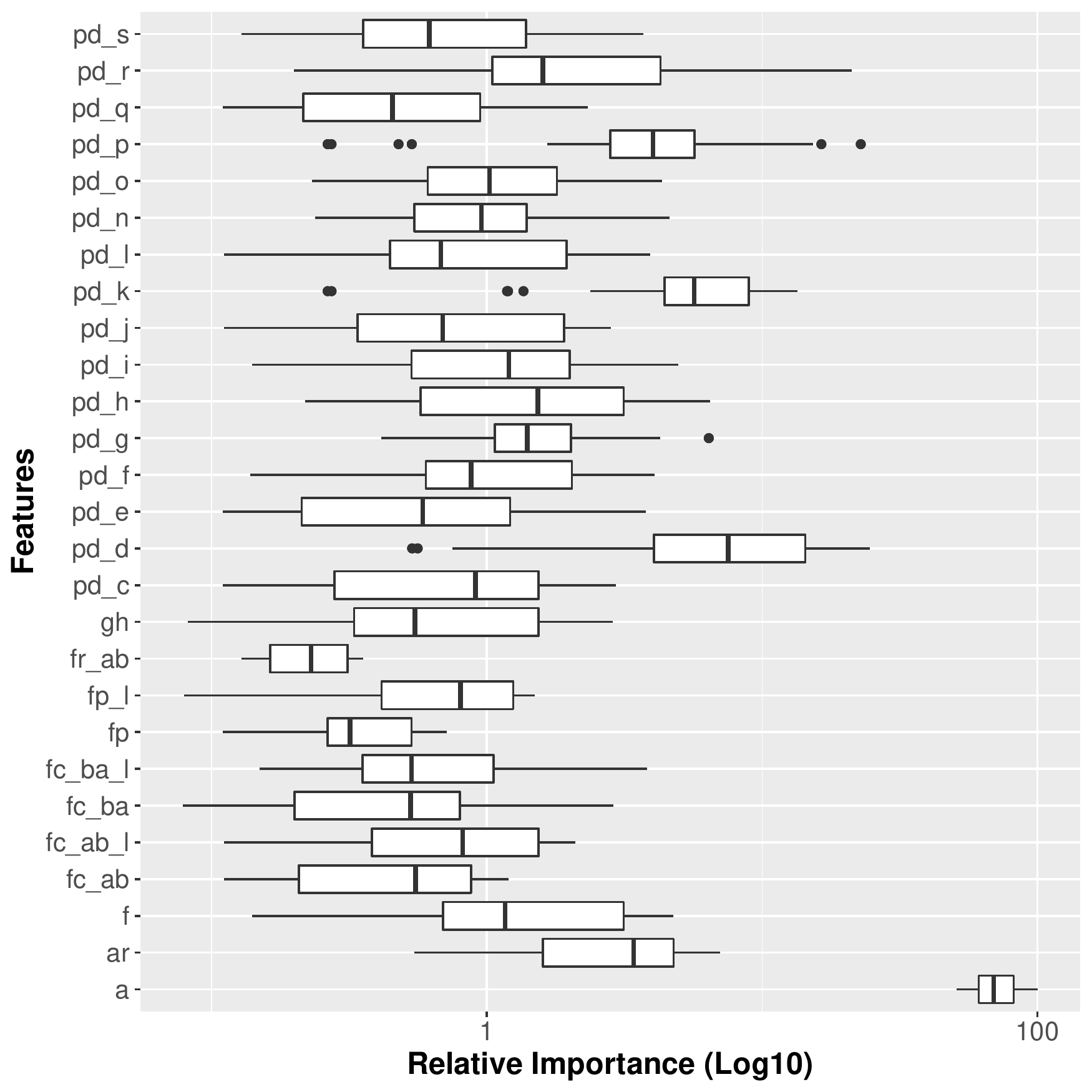}
    \caption{Box plots of the relative importance of each Profile Data feature }
    \label{fig:featureImportance}
\end{figure}

\begin{figure}[h]
    \centering
    \includegraphics[width = \linewidth, height = 7cm]{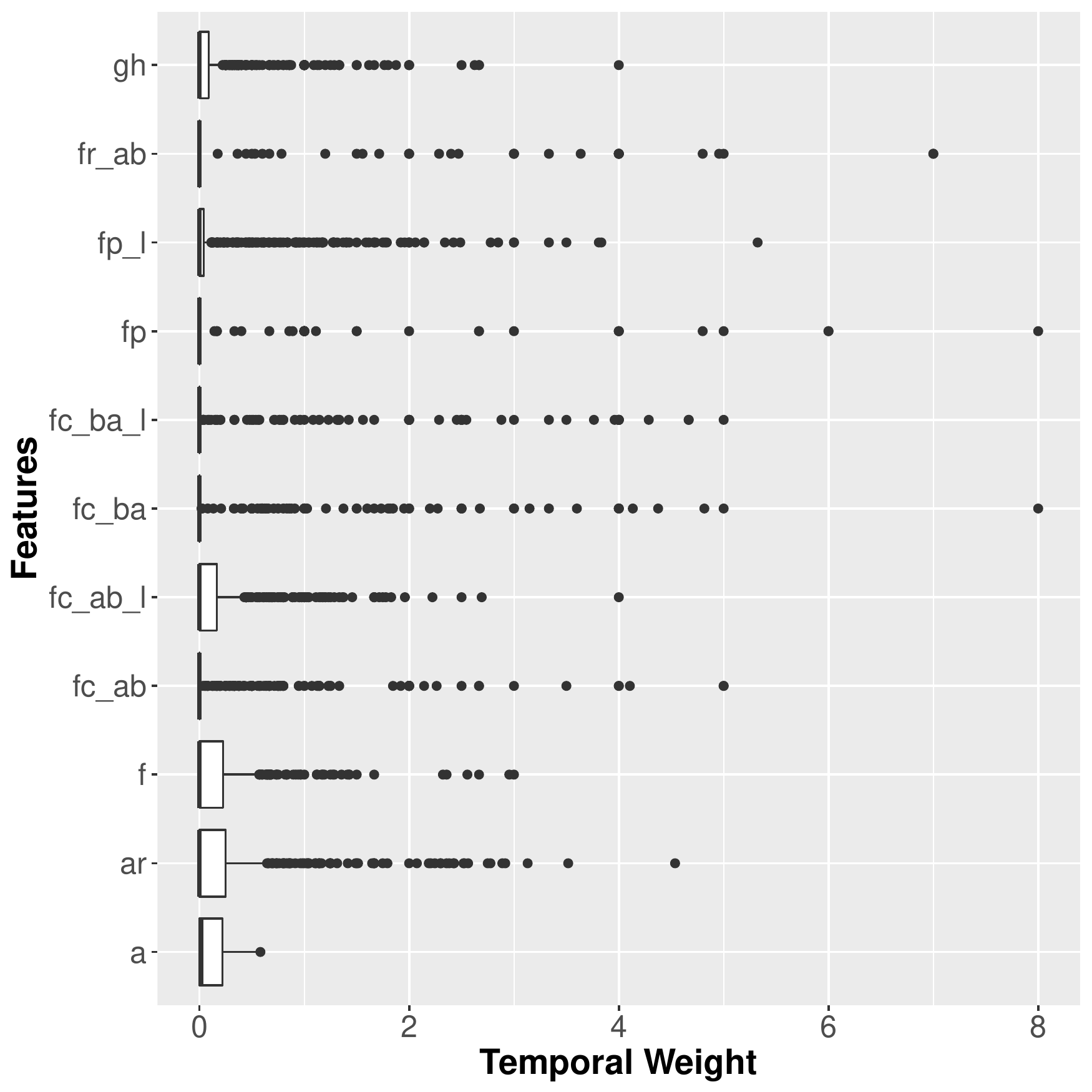}
    \caption{Box plots of the relative importance of each Profile Data feature }
    \label{fig:temporalWeightFeatures}
\end{figure}

To be able to judge of the importance of the different features, we set up an experiment similar to the previous one but we remove the users that did not fill their Profile. The reason for this is that having left the Profile empty is a very strong indicator of dropout and pushes the coefficients of the Profile features down. As mentioned previously, the number of temporal features increases as the lag increases, therefore, we chose to combine the values of different weeks of each feature while still taking into account the temporal quality that they possess. Figure \ref{fig:featureImportance} shows the relative importance (RI) of each feature on a logarithmic scale. These values are obtained from AdaBoost and are based on the mean decrease in the Gini Index. No pre-processing is done for the Profile Data features, however for the temporal features we sum the relative importance for the different values of each feature for each prediction. For example, if the relative importance for $a_0,a_1,a_2$ for week 2 are $0,15$ and $65$ respectively, then $a = 80$ for that prediction. We can see that 'a' is almost ten times more important than any other feature. Furthermore, we see that 'pd\_d','pd\_k','pd\_p' and 'ar' are relatively high compared to the other features.
To reflect how spread the weights of a certain temporal feature are over time, we define a new metric, the temporal weight, which is computed as follows:
\begin{equation}
    \textrm{TW}(x) = \sum_{i=0}^{lag} \frac{\textrm{RI}(x_i)}{\textrm{RI}(x)} \times (lag-i)
\end{equation}
Using the same values as the previous example, we would obtain: $\textrm{TW}(a) = \frac{0}{80} \times 2+\frac{15}{80} \times 1 +\frac{65}{80} \times 0 = \frac{15}{80}$. We then average these weights for each feature over all the predictions to obtain the box plots shown in Figure \ref{fig:temporalWeightFeatures}. We see that 'a', which has been shown to be the most important feature, has its values close to zero and only one outlier which implies that it mainly gives weight to the latest features. This can be explained by the fact that users either stop submitting assignments or persist until the end,but seldomly skip an assignment, hence looking more than one week back does not provide new information. The rest of these features denote activities which are not required to pass the course, which makes them interesting because they reflect a latent variable which can be defined as having an active attitude towards the MOOC. We see that all of them have many outliers and their range is significant, especially for 'fr\_ab','fp' and 'fc\_ba'. This implies that their values are often composed of the features of several weeks prior to the prediction. By analyzing the results of our experiments, we found that all these features are indicators of persistence rather than dropout. This knowledge combined with the results shown in Figure \ref{fig:temporalWeightFeatures} corroborates the hypothesis that Forum activity denotes an active attitude towards the course. Furthermore, the significant range of values suggests that the importance of these features barely decays over time, which means that having shown this active attitude towards the course at some point in time is still a strong indicator of persistence several weeks later.


\section{Conclusion}
\label{sect:discussion}
In this paper, a framework for extracting features from a MOOC course was presented and then these features were used as input to a classification system which is able to predict whether a user will dropout or not, taking into account the temporal dimension. In more detail, contributions of this paper are as follows: Firstly, predicting which users will drop out at some point is done with high accuracy for all three classifiers, even when using only the Profile Data. This is not to be translated that filling in the Profile leads to completing the course, but it does demonstrate that asking users to do so is important for teachers interested in predicting who will finish the MOOC.

Secondly, predicting the exact week that a user will drop out is considerably more difficult, and even when the lag is equal to the predicted week results obtained are not always looking good. 
This can be explained by the fact that there are few dropouts per week which makes it difficult for the classifiers to train and test, even when using k-CV. Furthermore, we believe that if the Video Data and the Team Data had temporal information, which would allow us to use them, better results could be obtained.

Finally, we found some interesting indicators of dropout and persistence. 
Although it is not a feature, we found that the users who leave their Profile empty have a high probability of dropping out, and this can be considered the strongest indicator of dropout. The strongest indicator of persistence is the assignment feature, which is the only feature that really denotes whether a user is actively involved in the course or not. Following this, we have a few features that reflect a hidden variable which could be to have an active attitude towards the MOOC. These features are the length of the Biography and of the Course objectives. Lastly, the temporal weights' graph  showed us that the features which denote an active participation of the user on the forum are relatively important. Given a dataset with more active forum users, it would be interesting to dig deeper into the relationship between Forum activity and user persistence.

Despite this work brought interesting insights into the dropout behaviour of users in a MOOC, there is still more work to be done. First of all, aligning all features so that they can be correctly projected on a temporal axis will boost the timely (per-week) prediction accuracy. Furthermore, current features can be combined with usage (log) statistics on the actual website (clicks, active time, etc.). Finally, we are looking forward to integrating the implemented approach with a real MOOC in order to check real-time performance and whether instructors are assisted in improving learners' experience.

\section*{Acknowledgment}
The authors would like to thank Daniëlle Verstegen (School of Health Professions Education, Maastricht University) for providing the dataset described in Section \ref{sect:dataset}.

\bibliographystyle{IEEEtran}
\bibliography{IEEEabrv,references}

\begin{thebibliography}{10}
\providecommand{\url}[1]{#1}
\csname url@samestyle\endcsname
\providecommand{\newblock}{\relax}
\providecommand{\bibinfo}[2]{#2}
\providecommand{\BIBentrySTDinterwordspacing}{\spaceskip=0pt\relax}
\providecommand{\BIBentryALTinterwordstretchfactor}{4}
\providecommand{\BIBentryALTinterwordspacing}{\spaceskip=\fontdimen2\font plus
\BIBentryALTinterwordstretchfactor\fontdimen3\font minus
  \fontdimen4\font\relax}
\providecommand{\BIBforeignlanguage}[2]{{%
\expandafter\ifx\csname l@#1\endcsname\relax
\typeout{** WARNING: IEEEtran.bst: No hyphenation pattern has been}%
\typeout{** loaded for the language `#1'. Using the pattern for}%
\typeout{** the default language instead.}%
\else
\language=\csname l@#1\endcsname
\fi
#2}}
\providecommand{\BIBdecl}{\relax}
\BIBdecl

\bibitem{khalil2014moocs}
H.~Khalil and M.~Ebner, ``Moocs completion rates and possible methods to
  improve retention-a literature review,'' in \emph{World Conference on
  Educational Multimedia, Hypermedia and Telecommunications}, vol. 2014, no.~1,
  2014, pp. 1305--1313.

\bibitem{balakrishnan2013predicting}
G.~Balakrishnan and D.~Coetzee, ``Predicting student retention in massive open
  online courses using hidden markov models,'' \emph{Electrical Engineering and
  Computer Sciences University of California at Berkeley}, 2013.

\bibitem{onah2014dropout}
D.~F. Onah, J.~Sinclair, and R.~Boyatt, ``Dropout rates of massive open online
  courses: behavioural patterns,'' \emph{EDULEARN14 Proceedings}, pp.
  5825--5834, 2014.

\bibitem{halawa2014dropout}
S.~Halawa, D.~Greene, and J.~Mitchell, ``Dropout prediction in moocs using
  learner activity features,'' \emph{Experiences and best practices in and
  around MOOCs}, vol.~7, 2014.

\bibitem{ramesh2014learning}
A.~Ramesh, D.~Goldwasser, B.~Huang, H.~Daume~III, and L.~Getoor, ``Learning
  latent engagement patterns of students in online courses,'' in
  \emph{Twenty-Eighth AAAI Conference on Artificial Intelligence}, 2014.

\bibitem{taylor2014likely}
C.~Taylor, K.~Veeramachaneni, and U.-M. O'Reilly, ``Likely to stop? predicting
  stopout in massive open online courses,'' \emph{arXiv preprint
  arXiv:1408.3382}, 2014.

\bibitem{cox1958regression}
D.~R. Cox, ``The regression analysis of binary sequences,'' \emph{Journal of
  the Royal Statistical Society. Series B (Methodological)}, pp. 215--242,
  1958.

\bibitem{menard2002applied}
S.~Menard, \emph{Applied logistic regression analysis}.\hskip 1em plus 0.5em
  minus 0.4em\relax Sage, 2002, no. 106.

\bibitem{janecek2008relationship}
A.~Janecek, W.~Gansterer, M.~Demel, and G.~Ecker, ``On the relationship between
  feature selection and classification accuracy,'' in \emph{New Challenges for
  Feature Selection in Data Mining and Knowledge Discovery}, 2008, pp. 90--105.

\bibitem{zou2005regularization}
H.~Zou and T.~Hastie, ``Regularization and variable selection via the elastic
  net,'' \emph{Journal of the Royal Statistical Society: Series B (Statistical
  Methodology)}, vol.~67, no.~2, pp. 301--320, 2005.

\bibitem{svetnik2003random}
V.~Svetnik, A.~Liaw, C.~Tong, J.~C. Culberson, R.~P. Sheridan, and B.~P.
  Feuston, ``Random forest: a classification and regression tool for compound
  classification and qsar modeling,'' \emph{Journal of chemical information and
  computer sciences}, vol.~43, no.~6, pp. 1947--1958, 2003.

\bibitem{pal2005random}
M.~Pal, ``Random forest classifier for remote sensing classification,''
  \emph{International Journal of Remote Sensing}, vol.~26, no.~1, pp. 217--222,
  2005.

\bibitem{freund1995desicion}
Y.~Freund and R.~E. Schapire, ``A desicion-theoretic generalization of on-line
  learning and an application to boosting,'' in \emph{European conference on
  computational learning theory}.\hskip 1em plus 0.5em minus 0.4em\relax
  Springer, 1995, pp. 23--37.

\bibitem{kohavi1995study}
R.~Kohavi \emph{et~al.}, ``A study of cross-validation and bootstrap for
  accuracy estimation and model selection,'' in \emph{Ijcai}, vol.~14,
  no.~2.\hskip 1em plus 0.5em minus 0.4em\relax Stanford, CA, 1995, pp.
  1137--1145.

\bibitem{he2009learning}
H.~He and E.~A. Garcia, ``Learning from imbalanced data,'' \emph{IEEE
  Transactions on knowledge and data engineering}, vol.~21, no.~9, pp.
  1263--1284, 2009.

\bibitem{weiss2004mining}
G.~M. Weiss, ``Mining with rarity: a unifying framework,'' \emph{ACM Sigkdd
  Explorations Newsletter}, vol.~6, no.~1, pp. 7--19, 2004.

\bibitem{huang2005using}
J.~Huang and C.~X. Ling, ``Using auc and accuracy in evaluating learning
  algorithms,'' \emph{IEEE Transactions on knowledge and Data Engineering},
  vol.~17, no.~3, pp. 299--310, 2005.

\bibitem{davis2006relationship}
J.~Davis and M.~Goadrich, ``The relationship between precision-recall and roc
  curves,'' in \emph{Proceedings of the 23rd international conference on
  Machine learning}.\hskip 1em plus 0.5em minus 0.4em\relax ACM, 2006, pp.
  233--240.

\end{thebibliography}
\end{document}